# Parametric four-wave mixing using a single cw laser

E. Brekke[*] and L. Alderson

*Department of Physics, St. Norbert College, De Pere, WI 54115*
*Corresponding author: erik.brekke@snc.edu*

Four-wave mixing can be used to generate coherent output beams, with frequencies difficult to acquire in commercial lasers. Here a single narrow ECDL locked to the two photon 5s-5d transition in rubidium is combined with a tapered amplifier system to produce a high power cw beam at 778 nm and used to generate coherent light at 420 nm through parametric four-wave mixing. This process is analyzed in terms of the intensity and frequency of the incoming beam as well as the atomic density of the sample. The efficiency of the process is currently limited when on resonance due to the absorption of the 420 nm beam, and modifications should allow a significant increase in output power.

The nonlinear process of four-wave mixing gives a means of generating coherent photons in atomic samples, and has shown promise for a wide range of applications. The process of four-wave mixing has been used in quantum information processing [1], investigation of Rydberg states [2-4], to investigate correlated photons [5], and in frequency up conversion.

The process of frequency conversion is especially appealing using parametric four-wave mixing (PFWM), where two of the four photons are generated during the subsequent decay cascade. This allows for the application of fewer laser systems, but also requires very intense laser beams or strong coupling. To date this has been pursued using either a single pulsed laser detuned from the intermediate state [6-9], or multiple low power lasers on resonance with the intermediate state [10-14]. For continuous schemes, it has been shown that the generated light has a sub-Doppler spectrum [14-18], enhancing the appeal for using this light in future applications.

The alternative method presented here allows both a simplified experimental setup and a simple means of investigating the frequency and intensity characteristics of PFWM. Following recent developments in tapered amplifiers [19], we generate a cw laser with high power while maintaining a narrow linewidth. Here a single high intensity cw laser is used to produce PFWM fields in a vapor cell of rubidium atoms for the first time. This novel process results in the generation of coherent 420 nm light, and an analysis of the system suggests excellent scaling with input power.

The excitation scheme and a schematic of the experimental setup are shown in fig. 1. The master laser is a standard ECDL locked to the $5s_{1/2} \rightarrow 5d_{5/2}$ transition using two photon spectroscopy [20], giving a detuning of 1 THz from the $5p$ state. This is sent through a homebuilt tapered amplifier system, giving 1.5 W of continuous power in the 778 nm beam, with an $M^2$ value of ≈2. The laser is focused to a waist of 50 μm, giving laser intensities up to $3.8 \times 10^4$ W/cm$^2$. The two photon Rabi frequency, $\Omega = \frac{\Omega_1 \Omega_2}{2\Delta}$, to the $5d$ state is 36 MHz. The input beam is circularly polarized to optimize the output power.

The amplified 778 nm beam is focused in a 5 cm long $^{87}$Rb vapor cell. This cell can be heated to temperatures up to 220 $^0$C, giving rubidium densities up to $2 \times 10^{15}$ cm$^{-3}$. In contrast to step-wise excitation, there is no precise spatial overlap of beams required for this process.

Here both of the first two photons in the four-wave mixing process are from the same laser. To satisfy phase matching, the resulting decay photons will be co-propagating with the original beam. A dichroic mirror and a bandpass filter allow the isolation of the 420 nm beam. The intensity of the 420 nm beam is monitored with a standard photodiode. There is not currently any means of monitoring the 5.23 μm beam.

The excitation process is far detuned from the intermediate state, so high intensity and atomic densities are necessary. Following the model for gain using the optical Bloch equations [7], we can determine the gain of the four-wave mixing process far from saturation,

$$g_{PM} = \frac{4\pi\eta |D_{5d,6p}||D_{6p,5s}|\Omega_2}{c\hbar |\Delta_{5d}||\Delta_{6p}|}\left[\frac{\omega_3\omega_4}{n(\omega_3)n(\omega_4)}\right]^{1/2}$$

where $\eta$ is the density of atoms, $\Omega_2$ the two photon Rabi frequency, $\Delta$ the detuning, $|D|$ the dipole matrix element, $\omega_{3,4}$ the frequency of the 5.23 µm or 420 nm beam respectively, and $n(\omega)$ the index of refraction.

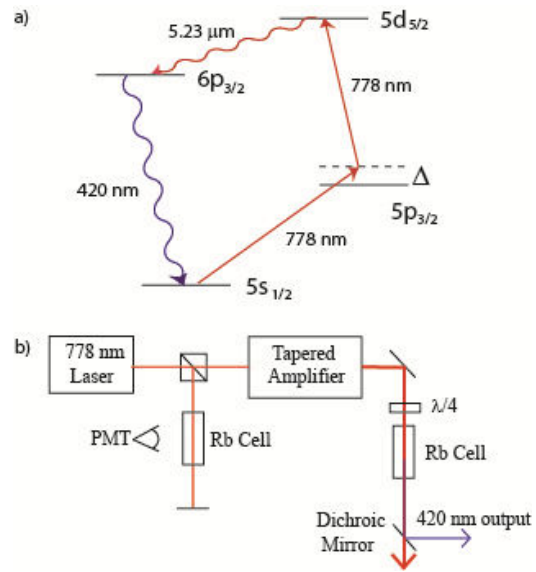

Fig 1. (Color Online) a) Energy levels for the PFWM process. b) A schematic diagram of the setup.

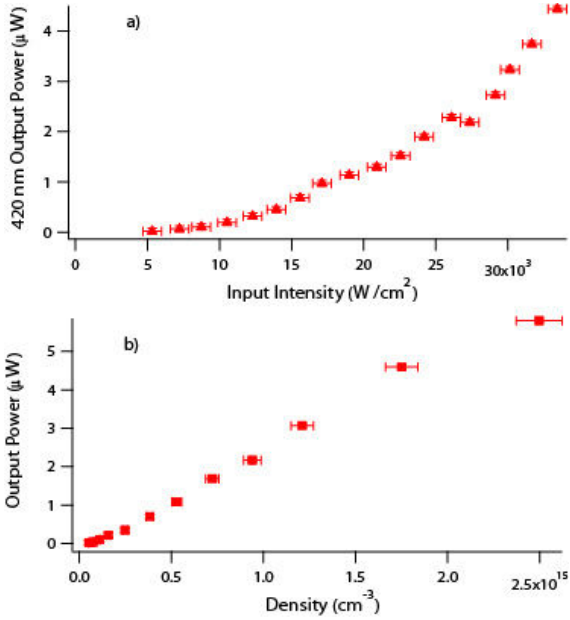

Fig 2. (Color Online) a) Output power at 420 nm vs input power. b) Output power at 420 nm vs atomic density.

For low densities and laser powers it is expected that maximum FWM gain will occur with resonant excitation and with the 420 nm beam on resonance with $6p\rightarrow 5s$, which has been observed. When the four-wave mixing process approaches saturation, maximum efficiency is expected with the 420 nm light detuned from the $6p$ resonance. For the parameters used here however, the process should remain far from saturation.

If the resulting 420 nm light is on resonance with the $6p\rightarrow 5s$ transition, high atomic densities can lead to rapid absorption. In order for the PFWM process to proceed, the gain must be larger than this absorption. For the data presented here, the location of the excitation beamwaist is adjusted to optimize output power. For an atomic density of $1\times 10^{15}$ cm$^{-3}$ we have observed a threshold intensity of $2\times 10^4$ W/cm$^2$ when on resonance with the $5s_{1/2}F=2\rightarrow 5d_{5/2}$ transition. Fig. 2 shows the power in the 420 nm beam as a function of the input power for an atomic density of $1.7\times 10^{15}$ cm$^{-3}$ and as a function of the atomic density at a laser intensity of $3.3\times 10^4$ W/cm$^2$. Increasing the atomic density also increases the absorption of the resulting 420 nm beam, and so gives limited increase in the efficiency of the process. The dependence on density shows signs of saturation, and gives approximately linear dependence. Increasing the intensity of the 778 nm laser, however, gives a rapid increase in the output power.

The frequency of the 778 nm laser can be scanned across both of the ground state hyperfine levels. The generated 420 nm light is shown as a function of the excitation frequency in fig. 3 for a variety of atomic densities and input laser powers. As the density of the sample increases, a doublet structure is seen, as the production of 420 nm light on resonance is suppressed. In contrast, as the 778 nm laser power is increased, the structure of the 420 nm light remains similar while the power output increases. The doublet structure seen here has also been seen in cases of two step excitation [11,21], but the mechanism was not fully understood.

In order to further investigate this structure, we translated the vapor cell along the direction of laser propagation while keeping the waist location constant. The beam's expansion over the length of the 5 cm cell results in the majority of the gain over a small portion of the cell. It is observed that when the waist is near the input side of the cell this doublet structure is pronounced, and approaches a Gaussian shape when near the output side of the cell. We propose that this structure is then a result of absorption of the 420 nm beam, increasingly noticeable at high atomic densities and optical depths.

The doublet structure suggests the frequency of the generated blue light depends on the excitation frequency. If the blue light produced off resonance is farther from the $6p\rightarrow 5s$ resonance, absorption would play a smaller role. The exact frequency of the blue light is determined by optimal gain of the four-wave mixing process, with complex dependence on phase-matching, detunings, refractive indexes, and intensities. However, it has been observed in the two-step excitation that the absolute frequency of the blue light follows the detuning from the 5$d$ state [21]. This experiment does not currently have a means of measuring the absolute blue frequency, but the structure displayed in fig. 3 suggests that detuned excitation results in the 420 nm light off resonant from the 6$p$→5$s$ transition. Since the resulting detuning is on the order of the Doppler width, this detuning would then lead to reduced absorption and higher efficiency blue light generation at high densities. This absorption effect has varying importance for the two hyperfine ground states, as a higher matrix element for the transition from the F=2 state results in higher absorption along this path, as can be seen in the structure of the peaks.

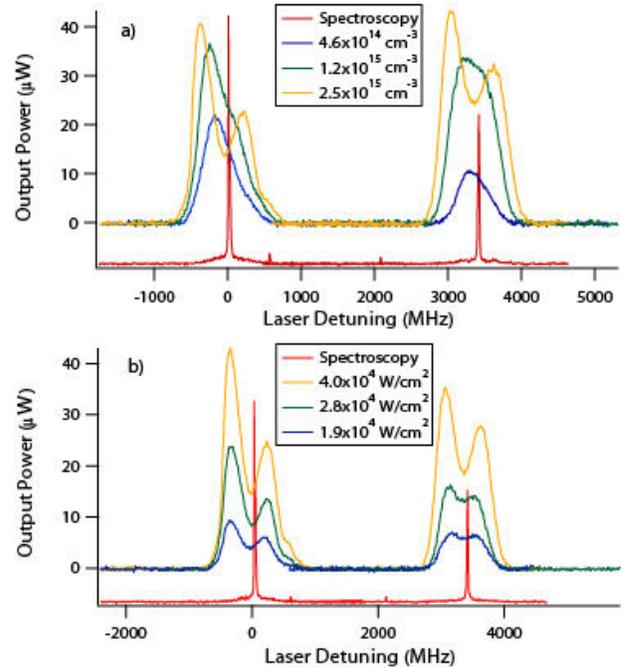

Fig.3.(Color Online) Output power at 420 nm vs excitation frequency for a) varying atomic densities with I=$4.0\times 10^4$ W/cm$^2$ b) varying laser intensities with $\eta$=$2.5\times 10^{15}$ cm$^{-3}$.

While increasing the atomic density will result in limited gains for the power output, rapid gain is seen as the excitation intensity is increased. Figures 2 and 3 show rapid growth, and the process is far from saturation. Significant increase in the output power has also been seen through the use of optical pumping of the 5$s$ hyperfine states [17]. This same technique could be used here to further increase the output power.

One of the great advantages of having a single frequency being used for both photons in the excitation process is the ease with which the intensity can be increased. A simple build-up cavity can be used surrounding the rubidium cell to increase the intensity.

Based on the gain observed in this process, this technique should be capable of generating several mW of 420 nm light when used with a ring cavity to build the laser intensity up to 25 times its current amount. While this output would fall well short of the best generated by frequency doubling a Ti:Sapphire laser [22] or commercial systems, it is comparable to lower cost alternatives [23] for similar input powers. The advantages of cheaper and simpler experimental setup make this process appealing for further investigation.

We have demonstrated the effectiveness of a novel means of coherent blue light generation using parametric four-wave mixing. A single ECDL and a tapered amplifier are needed, providing two-photon excitation while the remaining two photons are generated through the decay cascade. This process allows for a simple investigation of the frequency and power dependence. An investigation of the output power structure suggests that absorption plays a significant role, and light produced during off resonant excitation is detuned from the 6$p$→5$s$ resonance.

This four-wave mixing process presents an intriguing method for attaining novel wavelength lasers. Currently the efficiency of the process is limited by the absorption of the 420 nm beam in the optically thick rubidium cell when on resonance, giving output powers of 40 μW. The use of a build-up cavity to increase the intensity of the pump beam and optical pumping of the ground state should greatly increase the output possible.

In the future, the exact frequency of the 420 nm light can be studied to better understand the limiting efficiency of this process. We expect that the generated light should be both narrow and tunable near the 6$p$→5$s$ resonance.